\documentclass{iau}
\usepackage{graphicx}

\title[Extragalactic and Galactic jets] 
{Synergies in extragalactic and Galactic jet research}

\author[Gustavo E. Romero]   
{Gustavo E. Romero$^{1,2}$\thanks{Member of CONICET.}}

\affiliation{$^1$ Instituto Argentino de Radioastronom\'{\i}a (CCT La Plata, CONICET),\\ C.C.5, (1894) Villa Elisa, Buenos Aires, Argentina.\\ email: {\tt romero@iar-conicet.gov.ar} \\[\affilskip]
$^2$Facultad de Ciencias Astron\'omicas y Geof\'{\i}sicas, Universidad Nacional de La Plata, \\ Paseo del Bosque, B1900FWA La Plata, Argentina. \\email: {\tt romero@fcaglp.unlp.edu.ar}}

\pubyear{2015}
\volume{313}  
\pagerange{119--126}
\jname{Extragalactic Jets from Every Angle}
\editors{F. Massaro, et al.,  eds.}
\begin{document}

\maketitle

\begin{abstract}

 The discovery of relativistic jets and superluminal sources associated with 
accreting X-ray binaries in the Galaxy opened new ways of investigating the physics of 
outflows from compact objects. The short timescales and relatively large angular sizes 
of Galactic jets allow to probe the physics of relativistic outflows to unprecedented details. In this article I 
discuss results of recent modelling of Galactic jets, covering both radiative and 
dynamical aspects, which can shed light on different features of their extragalactic cousins.
 
\keywords{Jets, microquasars, active galactic nuclei}
\end{abstract}

\firstsection 
\section{Introduction}

The word `synergy' was introduced in English in the mid 19th century; it comes from the Greek `sunergos', meaning `working together', from `sun' -- `together' -- and `ergon', `work'. According to the Oxford's Dictionary the contemporary rendering is ``the interaction or cooperation of two or more agents to produce a combined effect greater than the sum of their separate effects.''

Research on relativistic jets has been dominated mostly by astronomical observations and theoretical studies of extragalactic jets. Only in the last 20 years relativistic jets have been identified in our galaxy (Mirabel and Rodr\'{\i}guez 1994, 1999). Since then an extensive monitoring of the radio and X-ray outflows of galactic binary systems has revealed many interesting features of nearby jets. A plethora of models has been developed to the point that in some cases the sophistication of theoretical representation of Galactic jets surpasses the more complex modelling of their bigger and more distant cousins. The current situation is such that fruitful exchanges can occur between Galactic and extragalactic jet researchers, with mutual benefits. The purpose of the following pages is to explore some topics particularly suitable for this synergy. I shall focus on aspects that I find specially interesting. There is no claim, hence, of completeness in my overview.   

\section{Jets: Galactic and extragalactic}

Galactic and extragalactic jets present many similarities, and also many differences. Both seem to be powered by accretion of magnetised matter with angular momentum onto a compact object; they both can display apparent superluminal motions; they seem to present a correlation of causal significance between the accretion and ejection processes; they display variability along the whole electromagnetic spectrum. These and other similarities are often stressed. The differences, however, are also important, and frequently overlooked. I shall mention the following ones:\\

\begin{itemize}  

\item Galactic jets seem to be slower. In some cases, where the bulk motions are directly measured through emission lines, the plasma velocity is $\sim 0.3 c$\footnote{Some very relativistic outflows have been reported, nonetheless. See Fender et al. (2004) for the most extreme case.}. 

\item Galactic jets  become dark not far from the central source. Typical lengths are $\sim 1000$ AU (e.g. Stirling et al. 2001).

\item They are heavy (have hadronic content) and produce thermal emission at their termination regions (Gallo et al. 2005, Heinz 2006).                                                                                                                                           

\item They interact with stellar winds (in the case of microquasars with high-mass donor stars; see Romero et al. 2003).

\item The size of the accretion disk is constrained by the presence of the donor star.

\item Gravitational and radiative effects of the donor star can be important to jet formation and propagation.

\item The disk and corona are hotter than in active galactic nuclei; magnetic fields are also higher close to the jet base.\\
\end{itemize}

These differences can yield, however, opportunities for observational and theoretical studies of physical situations that can be applied, with suitable modifications, to extragalactic jets. Before discussing such situations it is convenient to briefly review current models of microquasars's jets. 

\section{Models}

Models for Galactic jets in X-ray binary systems are based on original research of extragalactic jets, especially the pioneering models developed in the late 1970s and early 1980s by Blandford \& K\"onigl (1979), Marscher (1978, 1980), and Ghisellini  et al. (1985) among others. The situation has evolved to the point that while now one-zone or multi-zone models are still regularly used to explain the emission of extragalactic jets, Galactic jets are often represented using lepto-hadronic inhomogeneous models that include all kind of radiative and dynamical processes. Coupled transport equations are usually solved for all particle species, including transients mesons and muons (e.g. Bosch-Ramon et al. 2006; Romero \& Vila 2008; Reynoso \& Reynoso 2009; Vila et al. 2012; Zdziarski et al. 2012a, 2014a,b; Malzac 2014).

\begin{figure}[h]
 \vspace*{1.0 cm}
\begin{center}
 \includegraphics[width=3.4in]{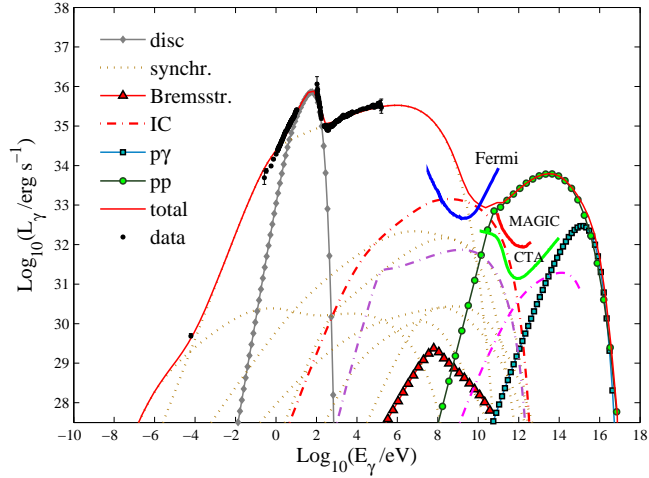} 
\vspace*{0.0 cm}
 \caption{Spectral energy distribution of a flare ocurred in 2000 in the low mass microquasar XTE J1118+480. The data have been fitted with an inhomogeneous letpo-hadronic jet model. For details see Vila et al. (2012). }
   \label{Romerog1}
\end{center}
\end{figure}

In most of these models physical conditions near the base of the jet are similar to those of the corona (e.g. Vieyro \& Romero 2012). Thermal plasma is injected close to the black hole and launched by magneto-centrifugal forces. Initially this plasma is magnetically dominated and mechanically incompressible. The plasma then accelerates longitudinally due to pressure gradients, and expands laterally at the sound speed (Bosch-Ramon et al. 2006). The gas cools as it moves outward along the jet. Extended inhomogeneous acceleration/radiation regions appear when shocks can finally be formed (Malzac 2014). A population of relativistic particles is then injected in the flow and the continuity equation for cooling and transport of the electron and proton populations is solved numerically. The result is a complex spectral energy distribution, with many components and some non-intuitive features arising from non-linear processes. In Figure \ref{Romerog1} I show an example computed by Vila et al. (2012) for a burst occurred in the low-mass microquasar XTE J1118+480 during 2000.

\section{Interactions}

The donor star in high-mass microquasars is an early-type star that ejects strong winds. The jet, originated close to the orbital plane, has to pave its way through the wind (Romero et al. 2003). Since the winds are clumpy (e.g. Owocki et al. 2009), jet-cloud interactions are expected to occur frequently (Araudo et al. 2009). Such interactions should result in gamma-ray variability as observed in sources such as Cygnus X-1 and Cygnus X-3 (e.g. Romero et al. 2010, Araudo et al. 2011, Piano et al. 2012). The gamma-ray radiation can be the result of either hadronic inelastic collisions between relativistic protons in the jet and nuclei in the wind or inverse Compton (IC) up-scattering of UV and X-ray photons locally produced in the shocked cloud. 

\begin{figure}[h]
 \vspace*{1.0 cm}
\begin{center}
 \includegraphics[width=5.50in]{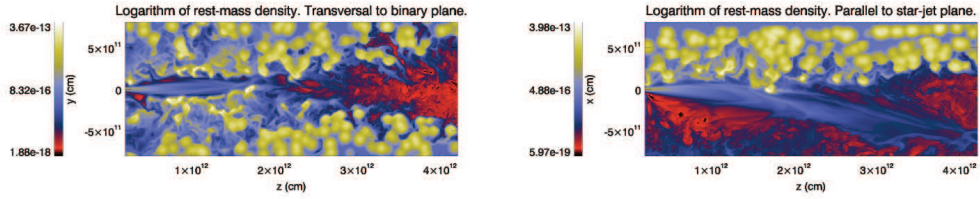} 
\vspace*{0.0 cm}
 \caption{Axial cuts of rest-mass density of a simulated 3D jet propagating through a clumpy wind in a high-mass microquasars. A clear bend is seen in the jet direction. From Perucho \& Bosch-Ramon (2012). }
   \label{Romerog2}
\end{center}
\end{figure}

\begin{figure}[h]
 \vspace*{1.0 cm}
\begin{center}
 \includegraphics[width=5.50in]{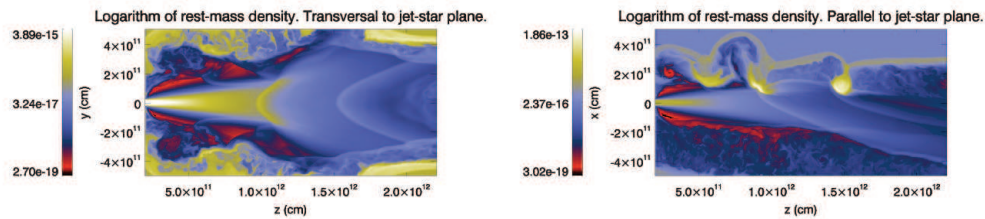} 
\vspace*{0.0 cm}
 \caption{Axial cuts of rest-mass density of a simulated 3D jet interacting with massive clumps in a high-mass microquasars. From Perucho \& Bosch-Ramon (2012). }
   \label{Romerog3}
\end{center}
\end{figure}

In addition to thermal and non-thermal radiation, strong perturbations in the jet flow are expected because of the wind-jet interaction. Recent simulations by Perucho \& Bosch-Ramon (2012) show that the jet can be highly distorted or even disrupted by the clumps. More homogeneous winds, impacting on the side of the jet, can result in large amplitude bends (see Figures. \ref{Romerog2} and \ref{Romerog3}). 

\begin{figure}[h]
 \vspace*{1.0 cm}
\begin{center}
 \includegraphics[width=3.0in]{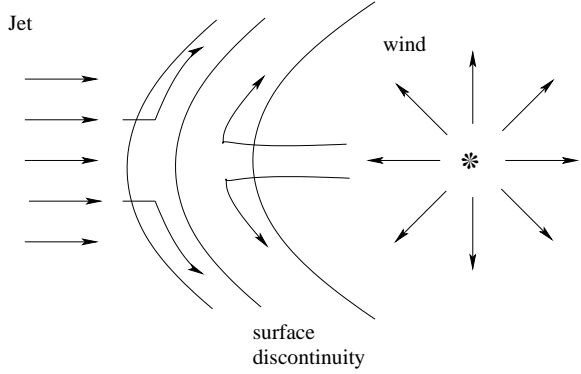} 
\vspace*{0.0 cm}
 \caption{Sektch of a massive star interacting with a relativistic jet in an AGN. See Araudo et al. (2013) for details.}
   \label{Romerog4}
\end{center}
\end{figure}

\begin{figure}[h]
 \vspace*{1.0 cm}
\begin{center}
 \includegraphics[width=3.4in, angle=270]{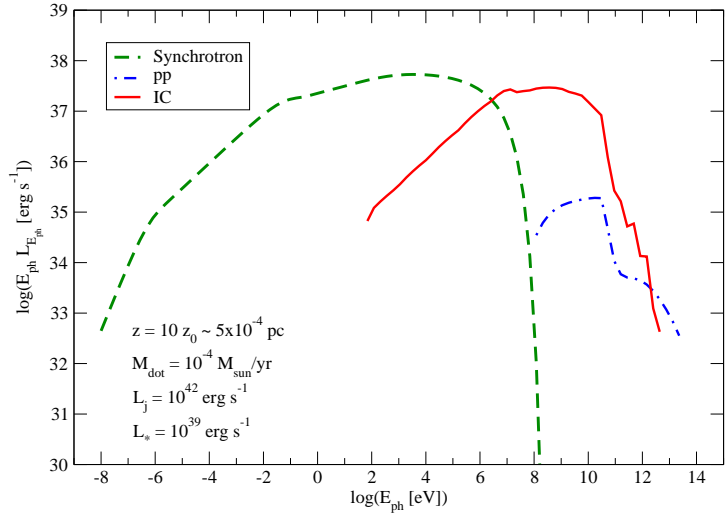} 
\vspace*{0.0 cm}
 \caption{Spectral energy distribution of the radiation produced by a Wolf-Rayet star interacting with a relativistic jet of a Faranoff-Riley I radio galaxy (Araudo et al. 2013).}
   \label{Romerog5}
\end{center}
\end{figure}

A similar situation to that of a clump interacting with a microquasar's jet can occur in an active galactic nuclei if the jet is penetrated by a massive cloud of the broad line region or by a star. In both cases, a bowshock will be formed around the obstacle. A reverse shock will move backwards the jet flow and a forward shock will propagate either through the cloud or the stellar wind. The cloud will be heated to X-ray emitting temperatures. This local photon field, then, will be up-scattered by reaccelerated electrons producing gamma-rays (Araudo et al. 2010, Bosch-Ramon et al. 2012). In the case that the interaction is with a supergiant star, the shock will blow away the stellar envelope producing a transient cloud (e.g. Khangulyan et al. 2013). If the star is of an early type, the stellar wind will be shocked and charged particles might be accelerated in both the wind and the shocked jet flow (see Fig. \ref{Romerog4} for a sketch of the situation). The stellar radiation field will provide copious photons for inverse Compton interactions (Bednarek \& Protheroe 1997).  Wind matter convected away through the bowshock can interact downstream the jet, producing gamma-rays through hadronic collisions with relativistic protons or heavy nuclei. Stars found by the jet on its way likely play an important role in the baryon loading of the flow (Komissarov 1994).  In Fig. \ref{Romerog5}, I show the transient spectral energy distribution of a flare caused by the penetration of a moderate-power ($\sim 10^{42}$ erg s$^{-1}$) jet by a Wolf-Rayet star.

\begin{figure}[h]
 \vspace*{1.0 cm}
\begin{center}
 \includegraphics[width=3.4in]{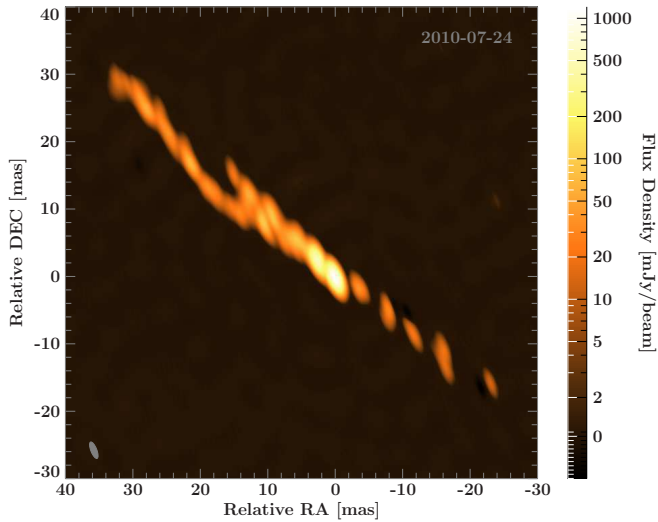} 
\vspace*{0.0 cm}
 \caption{TANAMI radio image of the inner jet of Centaurus A at 8.4 GHz. A ``tuning fork'' feature is clearly visible in the jet. It is likely caused by the interaction with a star or massive cloud. Scale is 1mas $\sim 0.018$ pc. From M\"uller et al. (2014).}
   \label{Romerog6}
\end{center}
\end{figure}

The interaction of jets and clouds should produce no only flares at high-energies, but also morphological perturbations in the jet. Very recently, M\"uller et al. (2014) have obtained a series of very high-resolution (sub-pc scales)  maps of the inner jet of the nearby FR I radio galaxy Cenaturus A. The images, taken from 2007 to 2011 at 8.4 GHz, show a peculiar feature resembling a ``tuning fork'', that seems to result from the interaction of the jet with a star or a massive cloud.  M\"uller et al. estimate the size of the bowshock to be $\sim 0.01$ pc, whereas the jet diameter at this point is of $\sim 0.1$ pc (see Fig. \ref{Romerog6}). Larger scale X-ray images show multiple acceleration sites along the jets (see Dan Harris's contribution to this volume). Such sites can be the bowshocks formed around different types of obstacles. Radio observations with high resolution will help to identify to nature of the obstacles, as shown by M\"uller et al. (2014). 

\section{Binarity}

The microquasar SS433 presents a classical example of a precessing jet (e.g. Fabrika  2004). 
The mass-loss rate in the jets is $\sim 5 \times 10^{-7}$ $M_{\odot}^{-1}$ yr$^{-1}$,
the period of precession is 162 d and the orbital period is 13.1 d. The jets are mildly relativistic, with bulk velocities of $\sim0.26c$. The precession is usually understood as the effect of the gravitational torque of the donor star on a misaligned accretion disk (Katz 1980, Larwood 1998). The disk precession is transmitted to the jets, producing a periodically variable emission (Reynoso et al. 2008a,b). The same effect is thought to operate in other microquasars (e.g. Kaufman Bernad\'o et al. 2002, Romero et al. 2002). Moreover, the same basic mechanism might work even in AGNs if a binary system of supermassive black holes is located at the center of a galaxy that underwent a recent merger. In case the accretion disk is not aligned with the orbital plane, the gravitational pull of the secondary black hole would trigger a precession that might modulate the superluminal ejections of plasmons (e.g. Abraham \& Romero 1999) and produce periodic variability on timescales of years (Romero et al 2000), in comparison with the much longer timescales of the Bardeen-Peterson effect (see Bicknell contribution to these proceedings). Several periods can be present in supermassive black hole binaries: those due to orbital interactions, gravitational perturbations on inclined disks, spin-spin precession, and precession caused by disk instabilities (e.g. Fan et al. 2001). 

A particularly interesting case of supermassive black hole binary is that of a system with a primary endowed with an accretion disk and a co-planar secondary moving inside the disk (Kocksis et al. 2012a,b). In such systems a gap can be opened in the disk by the secondary, and different accretion regimes are possible. In one of them, a matter inflow pervades the gap and feeds an inner accretion disk that can support a jet. The X-ray spectrum, nevertheless, is strongly disturbed from the standard case. This peculiar X-ray emission can be up-scattered by relativistic electrons and pairs at the base of the jet leading to specific high-energy features that can be used to probe the structure of the disk and as a diagnosis of close binarity in AGNs (Romero et al. 2014). Future high sensitivity gamma-ray instruments with pointing capabilities such as the Cherenkov Telescope Array might be used to detect these type of systems, which being in their last stage before the final merging would be potential targets for gravitational wave detectors.

\section{Cascades}

Electromagnetic cascades proceed wherever high-energy photons are injected in a region with a background photon energy density that exceeds the magnetic energy density. The gamma-ray photons are annihilated in the soft photon field producing electron-positron pairs. These pairs up scatter in turn soft photons into the gamma energy range, which results into a cascade that multiplies the number of pairs and redistribute the energy budget of the original photons. As a consequence, the spectral energy distribution can be strongly modified from the unabsorbed power law that initiated the cascade. If a strong magnetic field is present, the synchrotron cooling channel for the secondary pairs becomes dominant and the cascade can be cut off.  Cascades are then not expected close to the base of a jet that is magnetically launched. However, as the jet accelerates and expands, the magnetic energy density decreases and cascades can eventually develop. In high-mass microquasars the photon field of the donor star provides a suitable medium for gamma-ray absorption. The large opacities to gamma-ray photon propagation ensure efficient cascading in close binaries. Compact binaries with early-type stars such as Cygnus X-1 (Orellana et al. 2007 Romero et al. 2010, Bednarek \& Giovannelli 2007) and Cygnus X-3 (Cerutti et al. 2011, Zdziarski et al. 2012b) are very likely to sustain cascade production during gamma-ray flares.         

\begin{figure}[h]
 \vspace*{-6.0 cm}
\begin{center}
 \includegraphics[width=8.4in, angle=0]{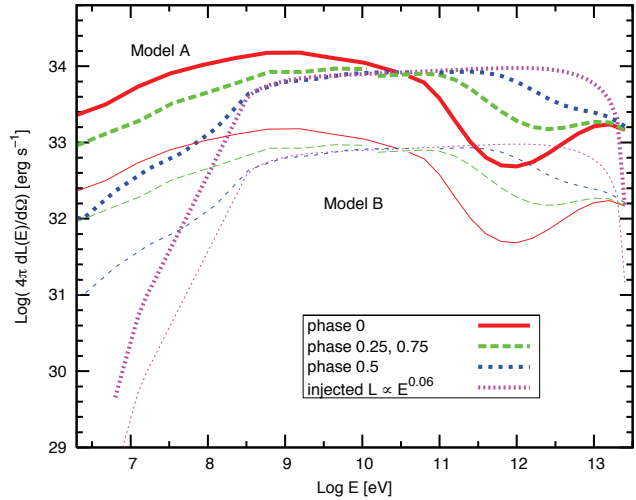} 
\vspace*{-3.0 cm}
 \caption{Spectral energy distributions resulting from electromagnetic cascades in the photon field of a massive star similar to the primary in Cygnus X-1. Spectral changes along the orbit are shown, for two different models of distinct inital power (Orellana et al. 2007).}
   \label{Romerog7}
\end{center}
\end{figure}

Typical cascade features in the UV field of an O-type star are shown in Fig. \ref{Romerog7}. Gamma-rays with energies in the range 0.1-10 TeV are efficiently absorbed and re-radiated at lower energies, producing a broad bump between 1 MeV and 1 GeV. Such bump might be detectable by future gamma-ray space detectors operating in the soft gamma-ray energy range (e.g. GAMMA-400, http://gamma400.lebedev.ru/indexeng.html).

The presence of weak magnetic fields produces a three-dimensionalisation of the cascade (Bednarek 1997, 2000, 2006; Pellizza et al. 2010). The pairs created can later diffuse and radiate at low frequencies forming radio halos around the sources (Bosch-Ramon \& Khangulyan 2011). Hadronic interactions close to the base of the jet can also trigger cascades by either secondary pairs or photons from pion decays (Orellana et al. 2007). 

In AGNs, infrared fields from dust tori and disk/jet photons reflected by clouds of the broad line region can provide a suitable medium for IC cascades. Hadronic-initiated cascades are also possible. Some authors have discussed cascading in the accretion disk photon field of AGNs (e.g.   Sitarek \& Bednarek 2010), but efficient multi-generation pair production in the vicinity of the supermassive black holes seems unlikely because of the large values expected for the magnetic field in these regions (Reynoso et al. 2011).   Interposed stars, on the contrary, can provide copious photons in low magnetic field environments.

\section{Final remarks}

Galactic and extragalactic relativistic jets present both similarities and differences. The main differences appear because of the presence of a donor star and the disimilar environment. But even the differences can be used to probe the relative incidence of similar physical processes. Because of the proximity of Galactic jets many physical processes that take place in mechanisms operating also in AGNs can be observationally probed in great detail. The use of VLBI radio observations to resolve jets during simultaneous X-ray and gamma-ray monitoring of microquasars remains to be fully explored. The implementation of this kind of multiwavelength studies would facilitate the construction of refined models that can be later extrapolated to AGNs and larger scales. Testing the new predictions of such models in a very different scenario would strengthen our understanding of the underlying physics. 

Many interesting aspects of the synergy between Galactic and extragalactic jets have not been discussed here. These aspects include the impact of the jets in the external medium, the disk-jet coupling, the matter content of the outflows, etc. Much can be learned from comparative studies of these topics.  

\begin{acknowledgements}
This work was supported by ANPCyT (PICT 2012-00878), Argentina, and MINECO under grant AYA2013-47447-C3-1-P, Spain. 
I thank Francesco Massaro, Teddy Cheung,  Ericson Lopez, and A. Siemiginowska for a delightful symposium. 

\end{acknowledgements}

\end{document}